\documentstyle[12pt]{article}
\input epsf
\begin{document}
\title{A determination of gluon spin distribution from the deep inelastic scattering data}
\author{ Jan
Bartelski\\ Institute of Theoretical Physics, Warsaw University,\\
Ho$\dot{z}$a 69, 00-681 Warsaw, Poland. \\ \\ \and Stanis\l aw
Tatur
\\ Nicolaus Copernicus Astronomical Center,\\ Polish Academy of
Sciences,\\ Bartycka 18, 00-716 Warsaw, Poland. \\ }
\date{}
\maketitle
\vspace{1cm}
\begin{abstract}
\noindent In order to determine polarized parton densities in
nucleon  we have made fits using all experimental data on spin
asymmetries measured in the deep inelastic scattering on different
nucleon targets. We have used in our analysis next to leading
order QCD corrections in $\overline{MS}$ renormalization scheme.
The functional forms of polarized densities are based on the fit
to unpolarized deep inelastic data made by the MRST group (MRST
99). We have concentrated on different models for gluon spin
distributions. In the best fit we get for gluon polarization (at
$ Q^{2}=1\, {\rm GeV^{2}}$)  $\Delta G=-1.0 \pm 0.6$. The obtained
result does not depend very much on various model assumptions.
\end{abstract}

\newpage

The experimental data on deep inelastic scattering of leptons on
nucleons enable us to determine parton densities in nucleons. For
polarized particles such experiments were made in SLAC
\cite{SLAC}, CERN \cite{CERN} and DESY \cite{DESY}. The results
were analyzed by many groups and several spin distributions and
different gluon polarization were presented
\cite{inni,gupta,BT,BTnn}. In this paper we do not consider any
new experimental data. All existing data on polarised deep
inelastic scattering were already  included in our latest fit
\cite{BTnn} where we have determined polarized parton densities
(performing next to leading order QCD analysis in $\overline{MS}$
renormalization scheme). However, since our method of
determination of spin distributions depends very strongly on the
parton distributions for unpolarized case we have performed a new
analysis using unpolarized distributions obtained by Martin,
Roberts, Stirling and Thorne (MRST) \cite{MRSTnew}.

The unpolarized densities are the sums of $+$ and $-$ components
(densities of quarks and gluons with helicity along or opposite to
the helicity of the parent nucleon) whereas the polarized ones are
given by the differences. The asymptotic behaviour of our
polarized parton distributions is determined (up to the condition
that the corresponding spin densities are integrable) by the fit
to unpolarized data.

The determination of unpolarized parton densities performed by
MRST \cite{MRSTnew} gives three solutions for gluon distribution.
In this paper we want to follow the method used in our previous
papers
 \cite{BT, BTnn} and to check how different functional form
 of gluon distributions influences our fits.
We will use data for spin asymmetries at given $x$ and different
$Q^2$. Altogether we will use in our analysis 431 experimental
points.  In our formulae for polarized densities we have included
12 parameters.

Experiments on unpolarized targets provide information on the spin averaged quark and gluon densities $q(x,Q^2)$ and
$G(x,Q^2)$ inside the nucleon. In ref.\cite{MRSTnew} these distributions (at $ Q^{2}=1\,{\rm GeV^{2}}$) are parametrized as follows:
\begin{equation}
q_{i}(x) = A_i x^{\lambda_i} (1-x)^{\eta_i} (1+\epsilon_i \sqrt{x}+\mu_i x),
\end{equation}
where one assumes that sea antiquarks and quarks of the same flavour have identical distributions.
The similar form one has for gluon density:
\begin{equation}
G(x) = B_G x^{\lambda_G} (1-x)^{\eta_G} (1+\epsilon_G \sqrt{x}+\mu_G x).
\end{equation}
The values of constants $\lambda_i$ and $\eta_i$, used in eqs. (1) and (2), are given in Table 1 for all three solutions which are called lower (L), central (C) and
higher (H) gluon solution in ref.\cite{MRSTnew} (the differences are visible for big $x$ values).
Our polarized densities for valence quarks (at $ Q^{2}=1\,{\rm GeV^{2}}$) are:
\begin{equation}
\Delta q_{i}(x) = x^{\lambda_i} (1-x)^{\eta_i} (\alpha_i+\beta_i \sqrt{x}+\gamma_i x),
\end{equation}
whereas for sea quarks and antiquarks one has:
\begin{equation}
\Delta \overline{q}_{j}(x) = x^{\lambda_s +\frac{1}{2}} (1-x)^{\eta_s} (\alpha_j+\beta_j\sqrt{x})+
c_i x^{\lambda_\delta} (1-x)^{\eta_s+2} \alpha_\delta (1+\epsilon_\delta \sqrt{x}+\mu_\delta x),
\end{equation}
where $c_i=+1 (-1)$ for up (down) flavour and zero for strange quarks. With such parametrization of quark spin densities we get from the fit
approximately $\alpha_\delta \approx 0$ and $\alpha_{\overline u} \approx \alpha_{\overline d} \approx 2 \alpha_{\overline s}$ ,
$\beta_{\overline u} \approx \beta_{\overline d}$.
Hence, our {\em{a posteriori}}  assumptions are:
\begin{eqnarray}
\alpha_\delta = 0, \nonumber \\
\alpha_{\overline u} =\alpha_{\overline d}=2 \alpha_{\overline s},\\
\beta_{\overline u}=\beta_{\overline d}. \nonumber
\end{eqnarray}
Such choice is consistent with the SU(2) symmetry in the sea.

For polarized gluon distribution we assume:
\begin{equation}
\Delta G(x) =  x^{\lambda_G} (1-x)^{\eta_G} (\alpha_G+\beta_G \sqrt{x}+\gamma_G x).
\end{equation}
The parameters $\alpha_i$, $\beta_i$ and $\gamma_i$
are determined from our fits and presented in Table 2.

\vspace{0.5cm}
Table 1. {\em {The values of constants which appear in eqs. (1-4,6) for three solutions of unpolarized gluon density. This figures determine the
behaviour of distributions at $x=0$ and $x=1$.}}

\vspace{0.3cm}

\begin{tabular}{||c|c|c|c||} \hline\hline
Constant&Lower &Central&Higher \\
&gluon solution&gluon solution&gluon solution
\\ \hline
$\lambda_{u_v}$&-0.527&-0.583&-0.560\\
$\eta_{u_v}$&3.44&3.43&3.45\\
$\lambda_{d_v}$&-0.748&-0.730&-0.740\\
$\eta_{d_v}$&3.86&3.86&3.95\\
$\lambda_{s}$&-1.240&-1.282&-1.273\\
$\eta_{s}$&7.63&7.65&8.30\\
$\lambda_{\delta}$&-0.356&0.183&0.157\\
\hline
$\lambda_{G}$&0.131&-0.031&-0.005\\
$\eta_{G}$&5.67&6.88&7.45
\\  \hline\hline
\end{tabular}
\vspace{1cm}

Table 2. {\em{The parameters of three fits calculated  at $Q^{2}=1\, {\rm GeV^{2}}$ .}}
\vspace{0.3cm}

\begin{tabular}{||c|c|c|c||} \hline\hline
Parameter&Lower &Central&Higher \\
&gluon solution&gluon solution&gluon solution
\\ \hline
$\alpha_{u_v}$&0.110&0.499&0.279\\
$\beta_{u_v}$&-3.50&-4.71&-4.22\\ $\gamma_{u_v}$&13.3&14.0&13.9\\
$\alpha_{d_v}$&-0.055&0.028&-0.059\\
$\beta_{d_v}$&-1.66&-1.70&-1.67\\
$\gamma_{d_v}$&0.007&-0.077&-0.007\\
$\alpha_{\overline{u}}$&-0.033&-0.133&-0.082\\
$\beta_{\overline{u}}$&1.54&1.37&1.58\\
$\beta_{\overline{s}}$&-0.402&-0.056&-0.226\\ \hline
$\alpha_{G}$&-18.3&-24.1&-20.3\\ $\beta_{G}$&52.5&117.2&86.7\\
$\gamma_{G}$&-48.8&-173.5&-127.9\\ \hline\hline
\end{tabular}
\vspace{0.5cm}

\noindent The total quark distributions are given by:
\begin{eqnarray}
\Delta u& = &\Delta u_{v}+2 \Delta \overline{u}, \nonumber \\
\Delta d& = &\Delta d_{v}+2 \Delta \overline{d},\\
\Delta s &=&2 \Delta \overline{s}, \nonumber
\end{eqnarray}
whereas one has for axial charges:
\begin{eqnarray}
\Delta \Sigma &= &\Delta u+\Delta d+\Delta s, \nonumber \\
a_{8} &=& \Delta u+\Delta d-2\Delta s, \\
a_{3} &\equiv& g_A= \Delta u-\Delta d. \nonumber
\end{eqnarray}

In Table 3 one can find above quantities integrated over variable $x$ and calculated at $Q^{2}=1\, {\rm GeV^{2}}$.
The axial charge $a_3$ is not fixed (usually one fixes it in an analysis) and comes out close to its experimental value.
The quantity $a_8$ is in some sense fixed i.e., it is included in a fit as an additional experimental point. We put
$a_8=0.58 \pm  0.1$, the value taken (with enchanced to 3$\sigma$ error) from the data on hyperon $\beta$ decays.

The typical errors of our determination are: $\Delta G = - 1.02 \pm 0.61$, $\Delta \Sigma =  0.04 \pm 0.18$, where we quote the figures for
higher gluon solution. We use all three MRST solutions keeping the quantity $a_8$ fixed. As it is seen from Table 3
the best $\chi^2$ we got for higher gluon solution.

\vspace{0.5cm}

Table 3. {\em{The integrated quantities from our three fits calculated  at
$Q^{2}=1\, {\rm GeV^{2}}$. }}
 \vspace{0.3cm}

\begin{tabular}{||c|c|c|c||} \hline\hline
Quantity&Lower &Central&Higher \\
&gluon solution&gluon solution&gluon solution
\\ \hline
$\Delta \Sigma$&0.19&0.02&0.04\\ $a_{8}$&0.58&0.59&0.51\\
$g_{A}$&1.25&1.29&1.31\\ $\Delta u$&0.79&0.75&0.75\\ $\Delta
d$&-0.46&-0.54&-0.55\\ $\Delta s$&-0.13&-0.19&-0.16\\ \hline
$\Delta G$&-0.77&-1.15&-1.02\\ \hline
$\chi^{2}$&363.91&363.16&362.28
\\  \hline\hline
\end{tabular}
\vspace{0.5cm}

For higher gluon solution we have tried to make several fits,
where we investigate other assumptions which slightly change gluon
distribution. In one of the fits we put $\alpha_{G_H}$=0 at
$Q^{2}=1\, {\rm GeV^{2}}$ (it means, that gluon density is less
divergent at $x=0$). In this case we get a fit which is of
undistinguishable quality from the best fit (in both cases
$\chi^2$ per degree of freedom is 0.86) and one gets $\Delta G$ =
- 0.63. However, when we put
$\alpha_{G_H}=\beta_{G_H}=\gamma_{G_H}=0$ (i.e. $\Delta G=0$ at
$Q^{2}=1\, {\rm GeV^{2}}$) we get $\chi^2$=366.0 (per degree of
freedom we have 0.87) which gives not much worse fit. We have also
performed other fits, one in which $a_8$  is treated as free
parameter and in another we allow gluon distributions $G^{+}(x) =
(G(x)+\Delta G(x))/2$ and $G^{-}(x) = (G(x)-\Delta G(x))/2$ not to
be positively defined. In both cases we have $\Delta G \cong -1.0$
and $\chi^2$ is nearly as good as in the best fit. In all
considered models we get negative polarization of gluons (the
negative value of gluon polarization one can find also in
\cite{gupta}). The values of gluon polarization (from our best
fit) at other $Q^2$ values are: $\Delta G (Q^{2}=5\, {\rm
GeV^{2}})=-2.0$ and $\Delta G (Q^{2}=100\, {\rm GeV^{2}})=-3.9$.
The variations of our model does not change significantly other
quantities (see Table 3).

In figures 1 and 2 we present the shapes of gluon spin distribution obtained in our fits at $Q^{2}=1\, {\rm GeV^{2}}$. One can see that the curves does not
differ very much.

 \begin{figure}
\noindent \epsfxsize =300pt\epsfbox{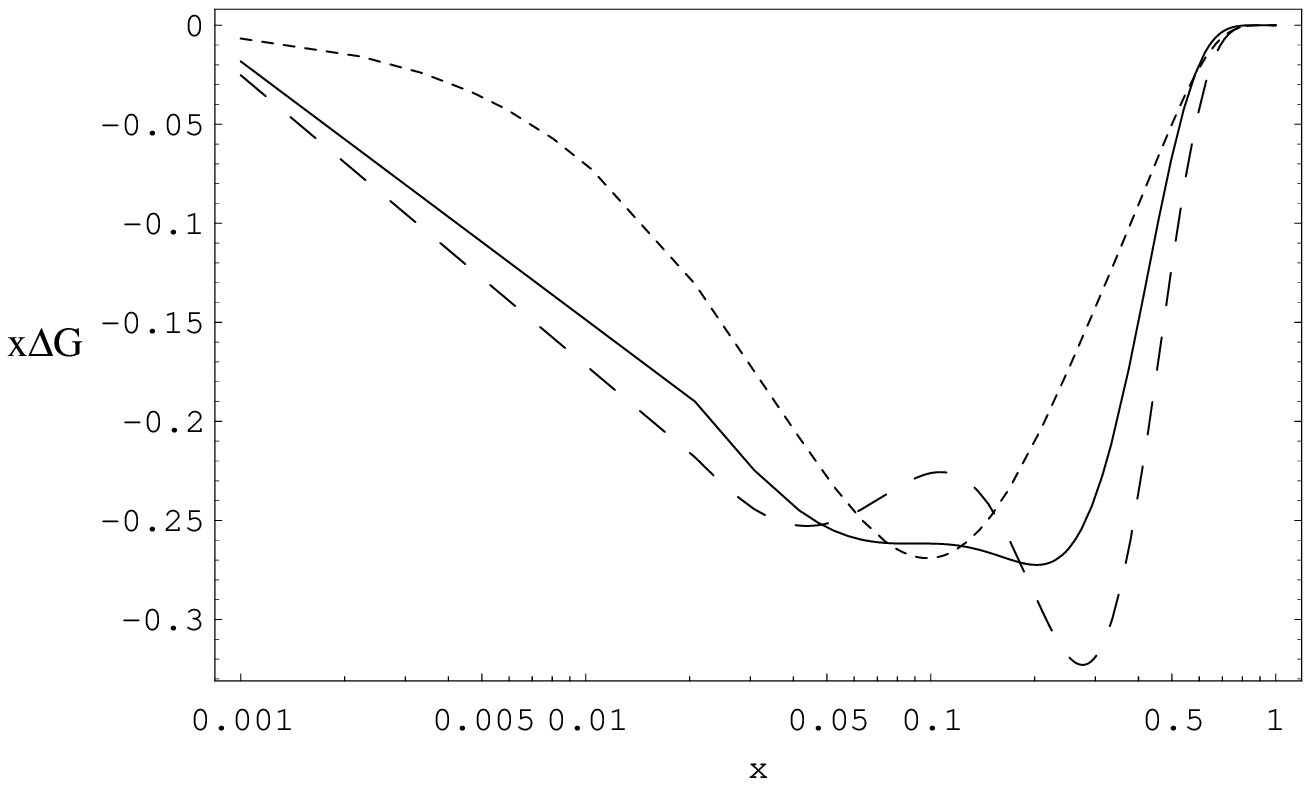} \vspace{0.5cm}
\caption{\em{The comparison of polarized gluon densities $x\Delta
G(x)$ versus $x$ obtained from the fits corresponding to lower
gluon (dotted line) central gluon (dashed line) and higher gluon
(solid line) solutions from ref.\cite{MRSTnew}.}}
\vspace{1.5cm}
\noindent  \epsfxsize =300pt\epsfbox{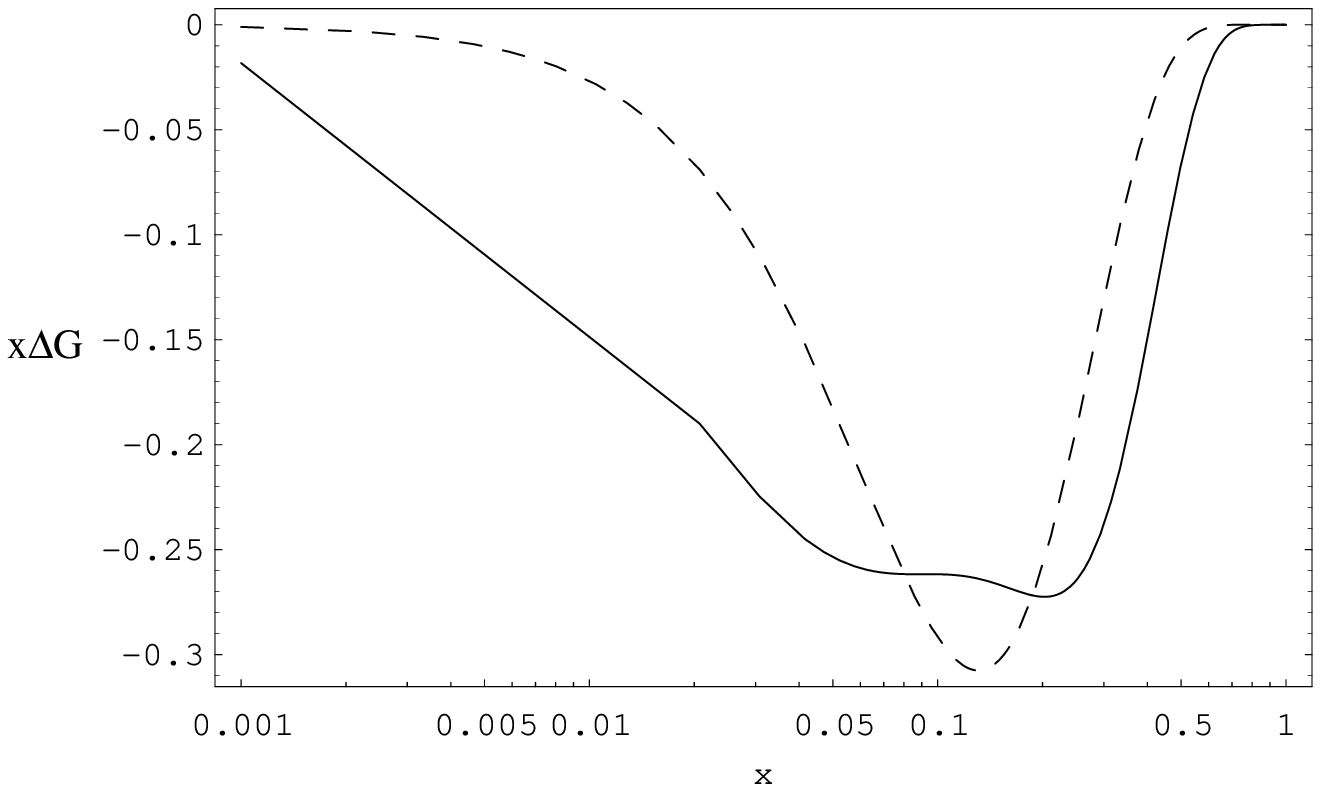} \vspace{0.5cm}
\caption{\em{The comparison of polarized gluon densities $x\Delta
G(x)$ versus $x$ obtained from the fits corresponding to the
higher gluon  (solid line) and the same model with
$\alpha_{G_H}=0$ (dashed line).}}
\end{figure}

We have made fits to precise data on spin asymmetries on proton, neutron
and deuteron targets. Our model for polarized parton distributions is
 based on MRST 1999 fit \cite{MRSTnew} to unpolarized data. Gluon
polarization comes out negative and relatively high. This conclusion does not depend very much
on different assumptions (within our famework) about gluon spin distribution.

\vspace{0.5cm}


\vspace{0.5cm}


\begin{thebibliography}{99}
\bibitem{SLAC} M.J.Alguard {\em et al.}, Phys. Rev. Lett. {\bf 37}, 1261
(1976); Phys. Rev. Lett. {\bf 41}, 70 (1978); G.Baum {\em et al.}, Phys. Rev. Lett. {\bf 45}, 2000 (1980);
{\bf 51}, 1135 (1983); E142 Collaboration, P.L.Anthony {\em et al.}, Phys. Rev.
Lett. {\bf 71}, 959 (1993); Phys. Rev. {\bf D 54}, 6620 (1996); E143 Collaboration, K.Abe {\em et al.}, Phys. Rev. Lett.
{\bf 74}, 346 (1995); E143 Collaboration, K.Abe {\em et al.}, Phys. Rev. Lett.
{\bf 75}, 25 (1995); E143 Collaboration, K.Abe {\em et al.}, Phys. Lett.
{\bf B 364}, 61 (1995); E154 Collaboration, K.Abe {\em et al.}, Phys. Rev. Lett.
{\bf 79}, 26 (1997); Phys. Lett. {\bf B 405}, 180 (1997);E143 Collaboration, K.Abe {\em et al.}, Phys. Rev.
{\bf D 58}, 112003 (1998); E155 Collaboration, P.L.Anthony {\em et al.}, Phys. Lett.
{\bf B 463}, 339 (1999); E155 Collaboration, P.L.Anthony {\em et al.}, Phys. Lett.
{\bf B 493}, 19 (2000);
\bibitem{CERN} European Muon Collaboration, J.Ashman {\em et al.},
Phys. Lett. {\bf B 206}, 364 (1988); Nucl. Phys. {\bf B 328}, 1 (1989); Spin Muon Collaboration, B.Adeva {\em et al.}, Phys. Lett.
{\bf B 302}, 533 (1993); Spin Muon Collaboration, D.Adams {\em et al.}, Phys. Lett.
{\bf B 329}, 399 (1994); D.Adams {\em et al.}, Phys. Lett. {\bf B 357}, 248 (1995);  B.Adeva {\em et al.}, Phys. Lett. {\bf B 412},
414 (1997); Spin Muon Collaboration, D.Adams {\em et al.}, Phys. Rev.
{\bf D 56}, 5330 (1997); Spin Muon Collaboration, D.Adams {\em et al.}, Phys. Lett.
{\bf B 396}, 338 (1997); Spin Muon Collaboration, B.Adeva {\em et al.}, Phys. Rev.
{\bf D 58}, 112001 (1998);
\bibitem{DESY} Hermes Collaboration, K.Ackerstaff {\em et al.},
Phys. Lett. {\bf B 404}, 383 (1997); Hermes Collaboration, A.Airapetian {\em et al.},
Phys. Lett. {\bf B 442}, 484 (1998);
\bibitem{inni} T.Gehrmann, W.J.Stirling, Phys. Rev. {\bf D53}, 6100
(1996); G.Altarelli, R.D.Ball, S.Forte, G.Ridolfi, Nucl. Phys.
 {\bf B 496}, 337 (1997); G.Altarelli, R.D.Ball, S.Forte, G.Ridolfi, Acta Phys. Pol.
 {\bf B 29}, 1145 (1998); Spin Muon Collaboration, B.Adeva {\em et al.}, Phys. Rev.
{\bf D 58}, 112002 (1998);  C.Bourrely, F.Buccella, O.Pisanti, P.Santorelli,
J.Soffer, Prog. Theor. Phys. {\bf 99}, 1017 (1998); E.Leader, A.V.Sidorov,
D.B.Stamenov, Int. J. Mod. Phys. {\bf A 13}, 5573 (1998);
E.Leader, A.V.Sidorov,
D.B.Stamenov, Phys. Rev. {\bf D58}, 114028 (1998);
Y. Goto et al. Phys. Rev. {\bf D62}, 034017 (2000);
D. de Florian, R. Sassot  Phys. Rev. {\bf D62}, 094025 (2000);
M.Gl\"{u}ck, E.Reya, M.Stratman, W.Vogelsang,
Phys. Rev. {\bf D 63}, 094005 (2001);
\bibitem{gupta} D.K.Ghosh, S.Gupta, D.Indumath, Phys. Rev. {\bf D 62}, 094012 (2000);
\bibitem{BT} S.Tatur, J.Bartelski, M.Kurzela; Acta Phys. Pol.
{\bf B 31}, 647 (2000); S.Tatur, J.Bartelski; Acta Phys. Pol. {\bf B 32}, 2101 (2001);
\bibitem{BTnn} J.Bartelski, S.Tatur; Phys. Rev. {\bf D 65}, 034002 (2002);
\bibitem{MRSTnew} A.D.Martin, W.J.Stirling, R.G.Roberts, R.S.Thorne;
Eur. Phys. J. {\bf C 14}, 133 (2000);

\end{thebibliography}
\end{document}